\documentclass{article}

\def\refs{\leftskip=.3truein\parindent=-.3truein}

\def\quote{\leftskip=.3truein\parindent=-.3truein\rightskip=.3truein}
\def\endquote{\leftskip=0.0truein\parindent=0pt\rightskip=0.0truein}

\begin {document}

\parskip = 3 mm
\parindent = 9 mm

\Large

\begin{center}
Where Did Copernicus Obtain the Tools to Build His Heliocentric Model? \\
Historical Considerations and a Guiding Translation of Valentin Rose's \\
``Ptolemaeus und die Schule von Toledo'' (1874)\\
\large
% [Version of 4 October 2017]
\end{center}

\large
\begin{center}
Kevin Krisciunas\\
Department of Physics and Astronomy, Texas A\&M University, College Station, 
Texas 77843 USA\\

and

Bel\'{e}n Bistu\'{e} \\
CONICET, Universidad Nacional de Cuyo, Mendoza, Argentina 
\end{center}

\Large
\begin{center}
Abstract
\end{center}

\large

We present a translation of the German text of an 1874 article by Valentin Rose 
that concerns the possible school of translators that worked in Toledo, Spain, from about 
1150 to 1250.  Rose's article relies significantly on the first-hand account of 
the activities in Toledo by the Englishman Daniel of Morley. The most prolific 
translator in Toledo was Gerard of Cremona, who translated Ptolemy's {\em 
Almagest} from Arabic into Latin with the help of Galib the Mozarab; this 
translation was significant to Copernicus's work.  Georg 
Peurbach and Regiomontanus based their {\em Epitome of the Almagest} (1463) on 
Gerard's translation, which in turn introduced Greek astronomy to astronomers in 
Italy and throughout Europe.  Copernicus studied in Padua in the first few years 
of the sixteenth century, where he learned about Ptolemy's {\em Almagest}.  
Copernicus's book {\em De Revolutionibus} (1543) also contains two geometrical 
tools of astronomers from thirteenth century Maragha, and his model of the motion 
of the Moon is mathematically identical to that of Ibn
al-Sha\hspace{-2 mm}$^-$t\hspace{-1.0 mm}$_{.}$ir (fourteenth
century Damascus). A Greek language manuscript written prior to 1308, and the 
residence in Padua of Moses Galeano, a Jewish scholar from Constantinople and 
Crete, who was familiar with the work of Ibn 
al-Sha\hspace{-2 mm}$^-$t\hspace{-1.0 mm}$_{.}$ir, provide evidence of the 
transmission of Arabic astronomical ideas to Copernicus.  We have only begun to 
understand this conduit of transmission.

\Large
\begin {center}
Part I. \\
A translation of (and guide to) Valentin Rose's \\
``Ptolemaeus und die Schule von Toledo'' (Ptolemy and the Toledo School)$^{1,2}$ \\
\end {center}
\large

Toledo was the place in which the threads of Platonic-Christian and 
Aristotelian-Arabic science were woven together during a minimum of a century 
(ca. 1150-1250), and especially during the correspondingly long reign of Alfonso 
VIII (1158-1214).  It was the seat of Christian power in Spain, having been 
conquered by Alfonso VI in 1085.  It was the capital of the Castilian region and 
the most important place in [Christian] Spain. For all Europe it was the hotbed 
of the ``doctrine of the Arabs.'' The colossal revolution of the era, the 
changed face of scientific activity, as if affected by a magic touch, [and] the 
fruitful zeal of the thirteenth century [shifted] to a previously unforseen 
field of work, by which a new selfstanding but comparatively humble spirit [of] 
the twelfth century [and its] applied offshoots evolved from [these] youthful 
foundations, such that everything was tied together in its origin in this city, 
in which, at the borders of the Arabic world and on the old foundations of 
Arabic education, the whole Western world was attracted with wonder to the 
evidence of this.  Here there were Arabic books in abundance as part of an 
[established] place of scientific scholarly activity [carried out by] a plethora 
of bilingual people.  With their help Arabo-Christians (Mozarabs) and 
long-settled Jews developed here a formal school of Arabic to Latin book and 
scientific manuscript translation, whereby those people eager to know science 
sought to learn Arabic and to participate in the work.  Numerous translations of 
the most famous writers of Arabic literature bear witness to this in Toledo.  
Englishmen and Germans, as well as Italians, linked the glory of their [careers] 
to their presence in this exalted school of Arabism and Arabic science.

The most productive of all translators who regularly brought new substance to the 
workmanship regarding all sciences, mathematics and astronomy, philosophy and 
medicine, Gerard of Cremona from Lombardy, spent almost his whole life in Toledo, 
learning and learning, translating and reading, reading before disciples of the whole 
world, who pursued the same purpose more and more here. He was effectively the father 
of translators (``who was the first among them'' said 
Roger Bacon).  That another famous Toledan translator, Heremannus Alemannus the Bishop 
[aka Hermann the German] was closely associated with him is convincing, according to 
the accounts of [their] contemporary Roger Bacon himself (opp. ined. I 471, see {\em 
Opus maius} p. 59), and even this Hermann also explains in his own lectures concerning 
Aristotelian books and concerning his ``Hispanic scholars''.  (There is the famous 
story of the Spanish word {\em belenum} [henbane], which could not be understood by 
the interpreters of Alfred's translation of Aristotle's {\em De vegetabilibus}, Rog. 
ib. I 467, see {\em Opus maius}, p. 45.)  According to his own admission (Rog. B. I 
472, see Jourdain {\em Rech.}, p. 140), Hermann translated with the help of Arabs, 
as did Michael Scotus. These four are the most famous translators in Toledo: Gerard of 
Cremona, Alvredus Anglicus [Alfred of England], Michael Scotus [Michael Scot], and 
Heremannus Alemannus [Hermann the German].  Thus, for example, in Toledo in 1217 
Michael Scot translated Alpetragius and then prior to ca. 1230 translated, with the 
help of one Andreas Iudaeus, Ar[istotle's] {\em De caelo et mundo} [On the Heavens] 
and other physical writings of Averroes [relating to] Aristotle, according to Roger, 
{\em Opus maius}, p. 36.  Also Hermann, who must have known Scot, tells us this 
(according to Roger I 472).

So, Hispanic scholars $-$ where?  In Paris?  Or in Spain itself?  In Toledo?

It is almost impossible to avoid saying, and there is according to my 
knowledge only medieval testimony for it, that Toledo at this time was 
virtually a kind of advanced school [{\em Art Hochschule}], if not a liberal 
arts school [{\em studium generale}] as somewhat later Alfonso VIII founded in 
Palencia [northwest Spain] on the basis of a Parisian model.  The scholars who 
came together here not only made books. First they themselves, then others, 
taught and gave open lessons. To convey this evidence is the purpose of the 
following short contribution.

It happens that there was an Englishman, Master Daniel, who the English 
bibliographers since Pits call Daniel of Morley (Morilegus of Leland).$^3$ 
Though there is more than one place called Morley, in this case it is Morley in 
the county of Norfolk, diocese of Norwich, because the man to whom [Daniel] 
dedicated his book was the Bishop John of Norwich, the well known John of 
Oxford, who served in that regard from 1175 to 1200.  In the preface he was 
called sacred mentor, and father.  All information about [Daniel's] life is 
based on this preface $-$ thus, virtually nothing is known about him. Of his 
books there is now only a single manuscript, and only this one did John Leland 
[1503-1552] often see, from which for its part Pits (Bale) and Tanner 
themselves established their own knowledge separate from those who later 
mentioned it (Wallis, Middeldorph, Jourdain, etc.) up to Thomas Wright, who 
first published it in the Biogr. Brit. lit. (II), London, 1846, p. 227-30.  It 
was a part of the preface, besides other short sections from the manuscript, 
namely from the Codex Arundel 377 (of the British Museum), membr. s. XIII 
(a well written small quarto).  The name of the author is shown here once in 
the title of the piece (folios 88-103) thus:

\begin{center}
The Philosophy of Master Daniel of M(or)lai \\
dedicated to Bishop John of Norwich
\end{center} 

Wright read directly and correctly in the name Merlai the undoubted variations 
(as did Leland, changing to Morley or Morlay and Merley) (according to the 
location of evident use in the same manuscript).  Thus, Daniel de Merlai.  In 
the preface he speaks of it, that he had been out of England for a long time, 
studying in Paris, but, soon unsatisfied with the honorable but scanty erudition 
[{\em Wissensschein}] of the teachers there, went to Spain in order to hear the 
``the wisest philosophers of the world'' (Wright, p. 228) at Toledo, the
famous center of Arabic science, ``which almost entirely appears in the quadrivium.'' 
With a rich collection of books he would return to England, where the liberal 
arts lay in a deep slumber.  Wright did not convey the rest of the preface: I 
shall make up for this, for he directly quotes the contents of the book, which, 
according to the manner of twelfth century philosophical representations of
Timaeus, carries the general title of philosophy, as the 
writings perhaps of Master Radulfus and Guilelmus de Conchis, whose widely 
circulated book (2nd edition: {\em secunda philosophia} or {\em maior}) follows 
a designated way in the same manuscript. It stands at the threshold of the two 
centuries: still on the dangerous floors of Platonic studies of creation, it is 
at the same time one of the first [examples] enfused with knowledge of the new 
Arabo-Aristotelian spirit, [based on] the adoption of Arabic astronomy and 
freely also [of] astrology. On account of his Arabs Daniel excuses himself in 
the preface but nevertheless this enthusiasm for it [astronomy] and its 
astrology come to be directly the reason that his book found no mercy in the 
eyes of the Church and was regarded as a poison.  He left hardly any track 
behind in the literature: no one read it and no one noticed.$^4$ Indeed the 
soon-to-be successfully demanding development of the literature may have rapidly 
relegated to the background the, for us, historically notable book, like so many 
other writings of the twelfth century (also that {\em Summa de philosophia} of 
Master Radulfus).

Literally, the most important thing is the conclusion: here Daniel speaks of his 
mentor in a long passage, which casts a certain bright light on the studies in 
Toledo, on the activivity of the translators and their whole operation [{\em 
Verfahren}].  Gerardus Toletanus held lectures for a circle of listeners (auditores), 
with whom he simultaneously debated [{\em disputiert}] (entirely in the Arabic 
manner, see Haneberg, {\em \"{U}ber das Schul- und Lehrwesen der Mohamedaner im 
Mittelalter} (Concerning the Learning and Teaching Methods of the Muslims in the 
Middle Ages), Munich, 1850, p. 24).  He gave lectures on astrology $-$ regarding that 
in a report about him there is the multi-layered [{\em eingeflochtene}, literally 
woven] statement ``qui Galippo mixtarabe interpretante almagesti latinavit'' [Gerard 
``rendered the Almagest into Latin with the Mozarab Galib as the one who was 
interpreting it.''$^5$] In connection to Ptolemy, his astrological writings became 
known through the work of Plato Tiburtinus [fl. 1116-1138, Italian mathematician, 
astronomer, and translator]. The translation of Ptolemy's {\em Syntaxis}$^6$ was 
certainly the most famous work of Gerard of Cremona, who, because of his long stay 
[in Toledo], is simply referred to as Toletanus. [Rose, p. 332, note 1, mentions a 
famous Spanish predecessor of the first half of that century, John of Seville $-$ 
John Toletanus (fl. 1136 to 1153).]  There exist many beautiful [manuscript copies] 
(and it was also published in Venice in 1515).  Out in front like a scholion 
[explanatory comment] is a kind of prologue, which Buttmann in the well known 
treatise about Claudius Ptolemy$^7$ spoke of and by means of its printing made it 
more widely known.  It arises in all likelihood (for it is said nowhere) from the 
translator and consists of a report about the Greek author, the Arabic translator and 
about the arrangement of the work.  The report on the author simply makes a 
contribution to the chapter regarding Ptolemy from the subsequent and also complete 
Latin translations of the known books of Abul Wefa (Albuguafe) [Abu al-Wafa, 940-998, 
Persian astronomer, worked in Baghdad].  The account of the author of the Arabic 
translation done under the auspices of 
al-Ma'mu\hspace{-2 mm}$^-$\hspace{-0.5 mm}n [ruled Baghdad 813-833 AD] reads 
according to the beautiful Wallerstein Manuscript [Codex Wallerstein, 1549], which I 
(while still in Mayhingen) once (1855) consulted myself (p. XIII, earlier in the 
possession of Michael Maestlin [1550-1631, mentor of Johannes Kepler] and Wilhelm 
Schikard): 

\quote
\parindent = 0 mm
\normalsize

%Liber hic precepto maimonis regis arabum qui regnavit in baldalt ablahazez [{\em 
%sic}] filio Josephi filii matre arismetici et sergio filio albe christiano in 
%anno duodecimo et ducentesimo secte saracenorum translatus est.

This book was translated, at the command of Maimon [al-Ma'mu\hspace{-2 
mm}$^-$\hspace{-0.5 mm}n], king of the Arabs, who reigned in Baldalt [Baghdad], by 
Al Abhazez son of Joseph, son of Matre the arithmetician [al-H\hspace{-2 
mm}$_{.}$\hspace{1 mm}ajja\hspace{-2 mm}$^-$j ibn \\
 Yu\hspace{-2 mm}$^-$suf ibn Matar], and [previously] by Sergio, son of 
Albe the Christian, in the year 212 of the sect of the Saracens.$^8$

\endquote
\large
\parindent = 9 mm

Usually in the manuscripts also follows afterwards the so called Commentary of Geber 
[Jabir ibn Hayyan, fl. 721-815, Persian chemist, astronomer, astrologer] (``Geber,
son of Afflah the Spaniard,'' according to the N\"{u}rnberger codex, see Bonc. Ger. p. 
13)$^9$ which Gerard had translated at the same time in Toledo.$^{10}$ A statement 
concerning the date of the translation of the {\em Syntaxis} (in the Florentine 
manuscript, Jourdain, p. 121)$^{11}$ is in fact without doubt, since the Arabic year 
(570) corresponds to the Christian date (1175) for the determination of the date of 
authorship of Daniel's book. This date is in the just-mentioned Ptolemy, Toledo 1175, 
and additionally we have the time period of 1175-1200 as the ruling period of the 
Bishop of Norwich to whom the book is dedicated.  Many things determine that Daniel 
was in Toledo between 1175 and 1187, probably soon after 1175.

[Here we paraphrase and elaborate the complex structure of the original.] 
Galen (129 to ca. 210 AD) was a 
Greek physician who was the authority on human anatomy until the publication 
in 1543 of {\em De humani corporis fabrica} by Andreas Vesalius.  Gerard's 
last translation was of Galen's {\em Tegni cum commento Haly} (``Tegni 
Galieni cum expositione ali abrodoan'' in the Index of Boncompagni, p. 6).  
Haly Abenrudian was Ali ibn Ridwan (ca. 988-1061).  At the end of Gerard's 
translation his ``socii'' (students, or collaborators) appended a brief 
obituary of Gerard, a list of his translations, and an eight line eulogy in 
verse.$^{12}$ The list is a commemoration of Galen's own list at the end of 
the very same work.  The ``socii'' lament that many of Gerard's translations 
are anonymous (i.e. not explicitly attributed to him) and after the death of 
their mentor they wanted to give credit where credit was due. [We revert to a 
more literal translation of Rose.] This index is for the most part completely 
ignored by the transcribers, often more or less mangled, but here is the 
place to find it. In the short vita dedicated to him it explicitly states 
that the man from Lombardy went to Toledo because of the {\em Almagest} 
(``amore tamen almagesti quem apud Latinos minime reperiit, Toletum perrexit'') 
[because of his love of the {\em Almagest}, which he did not find at all 
amongst the Latins, so he made his way to Toledo], and here in Toledo he 
remained ceaselessly translating until his death, which, according to a 
concurrence of information articulated in the manuscripts occurred when he 
reached 73 years old in 1187.  That Hermann the German actually collaborated 
with him [Gerard] in his [Hermann's] youth is difficult to substantiate.  
Hermann would have had to reach extremely old age [{\em steinalt} = as old as 
the hills], because in that case the passage of Roger [Bacon], which is as 
old as the whole book of the {\em Compendium studii philosophiae}, in which 
it is found, was written in 1271 (according to Brewer pref. p. 54), and he 
must have spent his whole life in Spain or returned there.  Hermann 
translated in Toledo in 1240 and 1256 (Jourd. p. 144. 142).

Because of the difficulty translating the {\em Almagest} it was naturally not 
the first item of Gerard's activity but it {\em was} the foundation of the 
determined prime purpose of his activity.  We see him as a person coming up 
the learning curve [{\em als einen Lernenden}] $-$ [with] ``Galippo mixtarabe 
interpretante'' [Galib the Mozarab, interpreting]. The Mozarab [Arabized 
Christian] translates into Spanish, and Gerard into Latin according to this 
guidance,$^{13}$ as already before the Jewish scholar John (David) of 
Seville, reported, by order of the Archbishop Raymond of Toledo (1126-1150), 
for the Archdeacon Master Dominicus Gundisalvi (Dominicus Gundissalinus) 
(Jourd. p. 108 ff.) Later under Alfonso X (The Learned) Aegidius de Thebaldis 
from Parma in his translation of Ptolemy's {\em Quadripartitum} points out 
himself that this was the general rule.  (The Spanish word {\em bele\~{n}o} is 
rendered thusly in the translation of Alfred the Englishman.)  However, the 
relationship of the roles of both collaborators was often different at 
different times: in the translations of John, who understood and wrote Latin, 
the progress with the Arabic of Dominicus seems nominal and diminishing, less 
so later on the progress of Galippus with Gerard.  We have no further clue of 
it except this one.

[Commentary:] Menocal comments on the process of collaborative translation in 
Toledo: ``The common model was for a Jew to translate the Arabic text aloud 
into the shared Romance vernacular, Castilian, whereupon a Christian would 
take that oral version and write it out in Latin.''$^{14}$ Richard Lemay 
notes that Galib is not named as a collaborator on any of Gerard's other 
translations.  He writes: ``The assumption that translators usually worked in 
pairs is an undue extrapolation from the very scanty occurrences.''$^{15}$

Galippus (Galib) was a Spanish Christian [in an] Arabic region $-$ 
``Mixtarabes'' is the Latin translation of what the Arabs and the Spanish 
chronicles called Mozarabs (i.e. the Arabized, but not of Arabic lineage).  
It is based on a phonetic allusion and modification of the Arabic word (see 
Florez, Esp. Sagr. vol. III, p. 190 ff.)  Properly understood the term means 
Christians and Arabs mixed together.  Starting in former times it is the 
acknowledged way of describing the Arabic word (see Charles du Fresne, sieur 
du Cange, {\em Glossarium mediae et infimae Latinitatis}, on the Mozarabs). 
As is well established, the subjugated Christians under the control of the 
Arabs, at least until the time of the Moroccan conquerors [Almohads and 
Almoravids], were undisturbed, not just regarding their religion but also the 
condition of their churches, and for a long time [also] their civil 
institutions.  So they stayed, but more and more adopted the language and 
customs of the conquerors. Legally, they were, through religion, a people 
unto themselves (if not also in fact), like having gone through a portal 
[{\em Mauer}], but informally they were assimilated in the service of the 
state.  We see Jews and Christian scholars, even bishops, in the Arabic 
capitals.  The Castilian kings (Alfonso VI through X) were, after the 
establishment of a new kingdom in Toledo, far from the dark barbarity of the 
book- and people-burning Catholic conquerors [later] in Granada.  Shortly 
before the conquest of Toledo [by the Christians] it was a flowering garden 
of Arabic science, in particular mathematical-astronomical science.  (I need 
only refer to the Toledan Tables.)  It is quite natural to think that Toledo 
desired foreigners and inhabitants, in part by maintaining business with the 
other Arabs in southern Spain, [by hosting] Arabs, Jews, and Mozarabs 
themselves, and in part through the collections of rare books in their Arabic 
libraries, whose reputation was widely known and long maintained.$^{16}$ 
(John of Seville and the physician Alcoati illustrate the connection between 
the ``enemy'' capitals Seville and Toledo during the reign of Alfonso VII, 
1126-1158.)  Such a one, and as far as we know the first, was Gerard of 
Cremona.  As said, he was the father of the genuine school of translation of 
Arabic science. At the same time, as we now learn, he was a real teacher, [a]
member of a real school, a kind of quasi-university of world science, which 
functioned as a direct beneficiary of the Arabic period. This led to the 
first actual and theological university and perhaps, by contrast, it was the 
cause of it.  The very same Alfonso VIII, who saw the flowering of those 
Toledan studies, was the founder of the first Spanish university in Palencia. 
The exact year of the founding is unknown, as well as the lifespan of this 
school. It has disappeared without a trace, and it has left no other memory 
behind than the noted report that the holy Dominicus, the founder (1216) of 
the Order of Preachers [Dominicans] should have studied there.  Both of the 
unique foundings, and both eyewitnesses of the founding, without mentioning a 
specific time frame under Alfonso VIII make reference to it.  Both the 
account of the battle at Las Navas de Tolosa (1212) [when Christians defeated 
the Muslims], as well as in the long peacetime since the defeat at Alarcos 
[1195, won by the Muslims] I will cite here, because they are noteworthy in 
this regard because of the statement concerning the convening of foreign 
teachers in Spain. There was one Rodericus Toletanus (the well known 
Archbishop, 1210-1247) [who] finished his chronicle after the concluding 
account in 1243, [where] he says, VII, 34:

\quote
\parindent = 0 mm
\normalsize

%Sed ne fascis charismatum quae in eum a sancto spiritu
%confluxerunt, virtute aliqua fraudaretur, sapientes a Galliis et Italia
%convocavit, ut sapientiae disciplina a regno suo numquam abesset, et magistros
%omnium facultatum Pallentiae congregavit, quibus et magna stipendia est
%largitus, ut omni studium cupienti quasi manna aliquando in os influeret sapientia
%cuiuslibet facultatis.  et licet hoc fuit studium interruptum tamen per dei gratiam 
%adhuc durat.

But lest the abundance of gifts from the Holy Spirit that had been joined in him be 
defrauded, he summoned learned men from France and Italy, so that the teaching of
knowledge would never be absent from his kingdom, and in Palencia he brought
together teachers of all the disciplines, to whom he granted large stipends, so
that knowledge of any field could flow like mana in the mouth of all who were
desirous of learning. And even if this study was once interrupted, yet by God's
grace it still continues today.

\endquote
\large
\parindent = 9 mm

(Thus, 1243).  The other [bishop] Lucas Tudensis (Bishop of
Tuy, 1239-1249, [who] concludes his book with the conquest of Cordova in 1236)
completes this report [with] an essential point by mentioning the actual author,
Bishop Tellus of Palencia (ruled 1212-1246, but in fact {\em electus} since
1208) p. 109 (Hisp. illustr. vol. IV): 

\quote
\parindent = 0 mm
\normalsize
%Eo tempore rex Adefonsus vocavit
%magistros theologiae et aliarum artium liberalium et Palentiae scholas constituit,
%procurante reverendissimo et nobilissimo viro Tellione eiusdem civitatis episcopo,
%quia ut antiquitas refert semper ibi viguit scholastica sapientia, viguit et militia.

In that time, King Alfonso called teachers of theology and of the other liberal 
arts and established schools at Palencia, under the administration of Tello, a
most reverend and noble man, bishop of this city, and indeed, as antiquity
reports, scholastic knowledge always flourished there, together with the military
spirit.

\endquote
\large
\parindent = 9 mm

After a comparison of this time period with the death year of King Alfonso 
VIII (1214), however, the authorship can only fall in the years 1208-14 or 
1212-14, that is either closely after, or what we remarked above is more 
likely just before Las Navas, thus either ca. 1208-12 just about when the {\em 
electus} [selection] of the author took place, rather than when he became the 
official bishop in the year 1212. Here this seems appropriate as it relates to 
the war breaking out, [but] a clean result is not possible.  That the 
continued existence of the school (under Ferdinand III, 1217-1252) had, by the 
way, already experienced an interruption, before 1243, we hear from Rodericus.  
Also, later $-$ after 1243 $-$ difficulties must have resurfaced, because a 
Papal Bull of Urban IV from 1263 sought (through assurance of privileges of 
Paris) in vain to improve the moribund studium ({\em scientarum studium 
generale} he called it) (see de Pulgar, Hist. de Pal. vol. II, p. 279).  Ch. 
Jourdain, Index chartarum univers. Paris. 1862, p. 28 no. 200, with which 
every doubt concerning the concealment [{\em Beseitigung}] of authorship [{\em 
Urheberschaft}] is to be compared [with a] totally similar expression in the 
Bull of the same Pope no. 189 from 1262.)  The Salamanca bull written a short 
time later led to auspicious progress.

\parindent = 9 mm

In spite of the fact that previously it had never been seen, the Toledan school, 
as well as world science, left behind rather far reaching tracks. 
[Paraphrasing...] Not only was it appreciated that the ancient Greek works were 
preserved and translated by the Latins, such as the astronomy of Ptolemy and the 
philosophy of Aristotle.  This connected the medieval scholars, as with a bridge, 
to the original sources.  We may note Albategnius's [al-Battini's] summary of 
Ptolemy and Avicenna's of Aristotle, in the first period of the school, as a 
result of Plato [Tiburtinus] and John, along with the most famous and most useful 
writings of their Arabic admirers, collaborators, and translators. This work was 
done shortly before in Toledo, and almost simultaneously in Seville and southern 
Spain by thriving Arabic scholars, who focused on the writings of astronomers 
like Alpetragius [al-Bitruji], for example, who wrote after 1185. It happened 
through the efforts of Michael Scot prior to 1217 relating to the work of 
Averroes [Ibn Rushd, 1126-1198] (whose work in Seville on Aristotle belongs to 
the period around 1170, [and] the Commentary on Physics, from the year 1186). 
[Resuming a more literal translation...] With Averroes translated into Latin a 
storm spread out from Toledo, which swept through all of Christendom and which 
provoked a huge radical change, through which the dialectic culture of centuries 
was determined, as long as the spirits were mature enough to follow Roger's call, 
[but] he himself was not mature enough to think of it himself. However, the fame 
of the place itself and the name Toledo was established not on these studies with 
such a bad scent (Averroes).  The timid charm of mathematical studies, which 
first had come to be known, encircled these names until late times. All the 
secret arts in astrology and magic which remained inseparable from Arabic 
understanding were determined for it. This dichotomy [{\em Zweideutigkeit}] also 
seeped through into the fame of the translations. I need to recall Michael Scot.  
But also the old mathematician Johannes Hispalensis [John of Seville] earned his 
magic glory late.  When in the year 1329 a great earthquake occurred in Bohemia 
and Bavaria, a document was circulated that was an epistle of astronomical 
science, in which the horrors of this year were prophesied many years before.  It 
read:

\quote
\parindent = 0 mm
\normalsize

%Magister Ioannes Davidis Toletanus et omnes magistri eiusdem loci universis ad
%quos presens epistola pervenerit, salutem ac sancti spiritus gratiam et
%solamen.  Noveritis quod a. d. 1329 mense Septembri sole existente in libra
%convenient omnes planete insimul, et sol erit in cauda draconis, et fiet 
%signatio rerum ammirabilium et horrendarum ... [Translated from German:
%The description of this very
%thing on which the council of writers] (nos cum magistratibus et sodalibus et
%peritis et astrologis Toletanis) [More from German: to defend themselves 
% against the earthquake by 
%means of appropriate house designs...] et sciatis quod nobiscum convenerunt
%omnes philosophi astronomi Hyspanie Grecie et Arabie et Hebrei.  audivimus
%etiam quod Meathinus turrim construxit excellentis edificii eritque turris
%ad instar unius magni montis, et aufivimus a rege Cycilie que nobis periculosa
%videntur.  Datum in Tolleto a. d. M.CCC.XXII.

[Here Rose quotes his source, in Latin, but in the middle makes some comments in German.]

Master John David of Toledo and all the masters gathered in this place to those 
to whom the present letter shall arrive, good health and the grace of the Holy
Spirit, and comfort. Know that in 1329 A.D. in the month of September with the Sun
in Libra all the planets will come together at the same time, and the Sun will be
in the dragon's tail, and a sign of strange and horrific things will be shown ...
(Then follows a description of this very thing which the council of writers [asserted was
necessary] to defend themselves against the earthquake by means of appropriate house designs.)
We also heard that Meathinus built a tower of excellent construction, and this tower
will be like a large mountain, and we heard from the King of Sicily things that look
dangerous to us. Dated in Toledo 1322 A.D.

\endquote
\large
\parindent = 9 mm

This pseudo-epigram reminds one with its presentation of spirited gatherings of
philosophers of a wonderful book from Spain of about the same time period, which
cannot be used as an historical source, but which doubtless serves as a reflection
of possibility.  I mean the ``Philosophia'' of Virgilius Cordubensis published
by Gott[hold] Heine, allegedly translated from Arabic 
%%%``in civitate Toletana a[nno] d[omini] 1290.''$^{17}$ 
``in the city of Toledo in the year 1290 AD.''$^{17}$
Toledo also appears here as the residence of the famous
schools, as a seat of philosophy and inclusion (pp. 212, 241-42) of all possible
magic arts.  Right at the beginning it says:

\quote
\parindent = 0 mm
\normalsize

%Cum apud civitatem Toletanam essent studia instructa omnium artium per magnum 
%tempus et loca scholarum extra civitatem essent posita et signanter studium 
%philosophiae esset ibi generale, ad quod studium veniebant omnes philosophi 
%Toletani, qui numero erant duodecim, et omnes philosophi Cartaginenses et 
%Cordubenses et Hyspalenses et Marochitani et Cantabrienses et multi alii qui erant 
%ibi studentes de aliis partibus, cum cotidie in scholis suis disputarent 
%philosophice de omni re, sic disputatione paulatim devenerunt ad quaestiones 
%difficiles ...

In the city of Toledo studies of all the arts were taught for a long time; the 
schools were placed outside the city and clearly the study of philosophy there was
general, and for this study came all philosophers of Toledo, who were twelve in
number, and all philosophers of Carthage and Cordoba and Spain and Morocco and
Cambridge and many others who were there as students from other parts. Every day
in their schools they disputed philosophically about every matter, and from this
disputation gradually developed difficult questions...

\endquote
\large
\parindent = 9 mm

The fame of the mathematical-physical studies is the long-lasting and original
things in these recollections of the thirteenth and fourteenth centuries, and the 
{\em studium philosophiae generale} we can regard as definitive because of
the travel report of Master Daniel.  Concerning the questions: how was it
learned, and where was it learned? We have a sketchy answer related to all
medieval universities.  Even regarding this embryo of a university in Toledo
prior to its actual birth [literally, prior to full maturity], one might gladly
consider it as established.  One part of the answer concerning the time period
can freely and without hesitation quote from the shadowy [{\em verkappt}]
Virgilius.  In all schools of Spain the lectures took place from the beginning
of October until May.  But the ``locale'' of the school?  Where was it held?
Soon after the conquest and contrary to the agreements of Christian
{\em Cultus} the great mosque of Toledo was forcibly taken over and reestablished
as a cathedral.  As with the Arabic schools and libraries in the mosques,
so it was also in Christian Toledo that the activity of its translators
happening under the auspices of the archbishop of Toledo (``the school of
translators,'' as Jourdain calls it on p. 108 and after him Renan Av. p. 201)
was connected spatially with the church of the archbishop.  Ferdinand III's
reign (1217-1252) brought to Spain the brilliant stimulus of architecture.
He was the architect of the magnificent Gothic cathedrals of Burgos and Toledo.
The former was built starting in 1221, while the latter somewhat later (the
Spanish say 1226 or 1227: see the indefinite account of Roder[icus] Tolet[anus]
IX, 13).  The construction went on for centuries, but at the beginning it
was, as always, asserted with naive enthusiasm  that a prompt date of
completion would be achieved.  In Burgos the essential construction was finished
in 1238, and ``it was raised from day to day not without great admiration from men,''
%%%``de die in diem non sine grandi admiratione hominum ... exaltatur,''
said the Archbishop Roderich in the year 1243 from Toledo.  Here too
the main construction, the actual house of the church, was finished in 1240, with
the old chapels of uniform disposition [{\em Anlage}], which in part later
were torn down and magnificently rebuilt in a semicircle around the enlarged
apse and, as Guhl says (Zeitschrift, 1859, p. 504), provide an eyecatching
view of the ground plan of the dome, [and] which correspond to the original
location of the building. To these chapels of the first [phase of] construction
belongs the still surviving though restored (de los Rios, p. 67) capilla de la
Trinidad (Guhl [{\em Erbkamschen Zeitschrift f\"{u}r Bauwesen}, 1858 and 1859], p. 498)
and on it one obtains the first and only date of construction concerning the
church, which indicates the actual time of construction.  We owe this to
Hermann the translator.  Because in this chapel, [and] often in the cathedral
he finished his translation of Averroes's Aristotelian ethics (Paraphrase).
It was (at that time) the ``school locale''.$^{18}$ In the Florentine manuscript
of the fourteenth century (concerning which [see] Renan Av. 212 contrasted by Jourdain,
p. 144) it says at the end: 

\quote
\parindent = 0 mm
\normalsize

%Dixit translator: et ego complevi eius translationem ex arabico in latinum tertio
%die iovis mensis iunii anno ab incarnatione MCCXL.  apud urbem
%Toletanum in capella Sanctae Trinitatis, unde sit domini nomen benedictum.

The translator said: And I completed this translation from Arabic into Latin on
Thursday, the third day of June, in the year of the Incarnation 1240, at the city
of Toledo, in the chapel of the Holy Trinity; for this praised be the name of the
Lord.

\endquote
\large
\parindent = 9 mm

So we see then from Raymond to Roderich, from ca. 1150 to 1250, there was a 
school of translators in connection with the Archbishopric and the Archbishop's 
church in Toledo.  There is no doubt that also the translations of Gerard of 
Cremona $-$ ``our Gerard, fountain, light, and glory of the clergy'' $-$ as well 
[as the translations] of ``philosophers of the world'' constituted a school of 
Arabic sciences, a ``liberal arts school,'' which functioned under the 
protection and under the roof of the old church, and which ``it was even 
standing, in the form of a mosque, at the time of the Arabs.''

\newpage
 
\Large
\begin {center}
Part II.\\
How Copernicus Became Familiar with  Ptolemy's {\em Almagest} and Ideas of the 
Maragha School
\end {center}

\large

Claudius Ptolemy wrote his {\em Almagest} in Greek in the middle of the second 
century A. D. The first Greek language printed edition was only published in 1538 (in 
Basel).  In the intervening centuries it served as the astronomical Bible thanks 
to Arabic and Latin versions.  The first Arabic translation was made by Sahl 
al-T\hspace{-2 mm}$_.$\hspace{0.5 mm}abar\i\hspace{-1.5 mm}$^-$, 
a Jewish scholar from Tabaristan (northern Iran, bordering the Caspian 
Sea).  More significantly, under the auspices of the Abassid Caliph 
al-Ma'mu\hspace{-2 mm}$^-$\hspace{-0.5 mm}n (who ruled in Baghdad from 813 to 833), 
al-H\hspace{-2 mm}$_{.}$\hspace{1 mm}ajja\hspace{-2 mm}$^-$j ibn Yu\hspace{-2 mm}$^-$suf
ibn Mat\hspace{-1.5 mm}$_.$\hspace{0.5 mm}ar (ca. 786-833) 
made his own Arabic translation in the late 820's based on a Syriac version of 
Sergios of Resaina (d. 536).$^{19}$ Later in the ninth century 
Ish\hspace{-1.5 mm}$_.$\hspace{0.5 mm}a\hspace{-2 mm}$^-$\hspace{-0.5 mm}q 
ibn H\hspace{-2 mm}$_{.}$\hspace{1 mm}unayn (ca. 879-890)
produced another Arabic translation, later revised by Tha\hspace{-2 mm}$^-$\hspace{-0.5 mm}bit
ibn Qurra (d. 901).  The translations of 
al-H\hspace{-2 mm}$_{.}$\hspace{1 mm}ajja\hspace{-2 mm}$^-$j and
Ish\hspace{-1.5 mm}$_.$\hspace{0.5 mm}a\hspace{-2 mm}$^-$\hspace{-0.5 mm}q-Tha\hspace{-2 mm}$^-$\hspace{-0.5 mm}bit
are still extant.$^{20}$

As one can see in the translation of Valentin Rose's 1874 article above, Gerard of 
Cremona was the most significant translator in medieval Toledo.  He worked there 
for more than 40 years, and his motivation for moving to al-Andalus was to seek 
out Ptolemy's {\em Almagest} and translate it into Latin.  In Rose's text and in 
the many footnotes he does not appear to give any details on which Arabic 
version(s) Gerard used, but it can logically be surmised that he used more than 
one.$^{21}$  As we know from a first hand source, the Englishman Daniel of Morley, Gerard 
was assisted in the {\em Almagest} translation by Galib the Mozarab, who rendered 
the Arabic into ``Romance vernacular,'' after which Gerard determined how best to 
express the same ideas in Latin.  It could have been Castilian, but others believe it
was the Mozarabic dialect.  There were other collaborative two-person teams 
of translators in al-Andalus in the twelfth and thirteenth centuries, but it is 
not certain that it was the {\em most common} practice to render a text orally 
into the vernacular, then write it out in Latin.  While the accepted date of the 
Gerard's translation of the {\em Almagest} is 1175, this is an upper limit based 
on a copy made by one Thaddeus of Hungary.  In any case it is a documented (and 
well accepted) fact that Gerard's translation was the prime conduit of ancient 
Greek geocentric astronomy to medieval and Renaissance Europe.  It was first 
printed in Venice in 1515.$^{22}$

The Polish astronomer Nicholas Copernicus (1473-1543) studied law, medicine, and 
astronomy in Italy (Bologna and Padua) during the years 1496 to 1503.$^{23}$  This is 
where he became familiar with the {\em Epitome of the Almagest}, started by Georg 
Peurbach (d. 1461),$^{24}$ and finished by Johannes M\"{u}ller (1436-1476),$^{25}$ 
known as Regiomontanus, about 1463.  (An ``epitome'' is not a summary {\em per 
se}.  It is a selection of quotations.)  Peurbach and Regiomontanus relied on 
Gerard's Latin translation.  The {\em Epitome} was printed in Venice in 1496. 
Copernicus became familiar with its contents.

The classic story of how Copernicus revolutionized planetary astronomy asserts 
that he relied on Euclidean geometry, Ptolemy's {\em Almagest}, and a big, bold 
insight.  But over the past 60 years we have come to realize that Copernicus's 
great book {\em On the Revolutions of the Heavenly Spheres} (1543) significantly 
uses two mathematical tools from thirteenth century astronomers of the Maragha 
Observatory (in Azerbaijan), and that his model of the motion of the Moon is
mathematically identical to that of a fourteenth century astronomer from
Damascus.  Much useful discussion and many useful references can be found in
George Saliba's excellent book {\em Islamic Science and the Making of the European
Renaissance}.$^{26}$

Now, the two fundamental axioms of ancient Greek astronomy were: 1) the Earth is 
the center of the cosmos; and 2) motion of celestial bodies (the Moon and all more 
distant objects) is characterized by {\em uniform} circular motion. To explain the 
retrograde motion of a planet or the varying angular speed of a planet, the Greeks 
invented certain geometrical devices to construct models of the motion of the Sun, 
Moon, and planets.  The primary orbit of an object around the Earth is called the 
{\em deferent}.  This governs the {\em direct} motion of a planet $-$ west to east 
against the stars. Retrograde motion was explained by placing the planet on an 
epicycle, whose center moves uniformly along the deferent. When the (faster) 
velocity of a planet along the epicyle is in the opposite direction to the 
direction that the deferent turns, an observer on the Earth observes the planet to 
exhibit retrograde motion (east to west) against the stars.  Here uniform 
motion means speed along some circle, in units of distance per unit time.

Ptolemy modified the first axiom listed above.  He suggested that the Earth could 
be offset from the exact center of the deferent of some orbiting body.  And he 
postulated a new point, the {\em equant}, which is situated on the other side of 
the center of that deferent at a distance from the center equal to the amount the 
Earth is offset from the center on its own side.  Furthermore, the center of a 
celestial body's epicycle is allowed to move uniformly, in {\em degrees per day}, 
with respect to the equant point.  Keep in mind that the motivation of a 
Ptolemaic model of the motion of a celestial body was to fit the measured 
coordinates of that object (i.e. in the past) and to predict the {\em direction} 
toward that object in the future.  The implied {\em distance} to a celestial 
object was of lesser concern.  For example, Ptolemy's model implied that the 
Moon's distance varies from 33.55 to 64.17 Earth radii, nearly a factor of two.  But 
our own naked eye observations of the angular size of the Moon (which varies inversely 
with the distance) show that the range of the Moon's distance is about $\pm$ 4 
percent.$^{27}$ A modern model of the Moon's motion shows that at perigee its 
distance is up to 7.3 percent less than the mean distance of 60.27 Earth radii,
while at apogee the distance is 5.8 percent greater than the mean distance.$^{28}$

The opinion of many historians is that Ptolemy's equant was very clever, 
but the medieval astronomers of the ``Maragha school'' objected to its use.  Uniform 
motion, measured in degrees per day with respect to the equant point, is not the 
same as uniform speed along a circle. According to Aristotelian physics motion in 
the terrestrial realm is rectilinear.  Water can flow downhill.  A rock falls 
toward the center of the Earth.  Air and fire can go up, in a line.  But motion in 
the celestial realm is supposed be along perfect circles.  So, if one needs an
oscillating linear offset to account for the motion of a planet, how can one achieve 
this by a combination of circles?  Such a method was invented by 
the director of the Maragha Observatory,
Nas\hspace{-1.5 mm}$_{.}$\hspace{0.5 mm}\i\hspace{-1.5 mm}$^-$\hspace{-1 mm}r 
al-D\i\hspace{-1.5 mm}$^-$\hspace{-1 mm}n
al-T\hspace{-2 mm}$_{.}$\hspace{1 mm}$\overline{\rm{u}}$s\i\hspace{-1.5 mm}$^-$ (1201-1274).
He showed how a 
smaller circle can turn inside another circle with twice the diameter, and a point 
on the diameter of the larger circle can oscillate back and forth linearly, even 
though it is produced through the combination of two uniform circular motions.  
This has been known as a 
``T\hspace{-2 mm}$_{.}$\hspace{1 mm}$\overline{\rm{u}}$s\i\hspace{-1.5 mm}$^-$
couple'' since 1960 when Edward S. Kennedy named it so.$^{29}$
For a diagram see Fig. 4.7 of Saliba,$^{30}$ or 
Nas\hspace{-1.5 mm}$_{.}$\hspace{0.5 mm}\i\hspace{-1.5 mm}$^-$\hspace{-1 mm}r 
al-D\i\hspace{-1.5 mm}$^-$\hspace{-1 mm}n's actual exposition.$^{31}$  
An animation can be seen at the {\em Wikipedia} page about 
Nas\hspace{-1.5 mm}$_{.}$\hspace{0.5 mm}\i\hspace{-1.5 mm}$^-$\hspace{-1 mm}r 
al-D\i\hspace{-1.5 mm}$^-$\hspace{-1 mm}n.$^{32}$

A second geometrical tool, attributed to 
Mu'ayyad al-D\i\hspace{-1.5 mm}$^-$\hspace{-1 mm}n
al-Urd\hspace{-1.5 mm}$_.$\hspace{0.5 mm}\i\hspace{-1.5 mm}$^-$ (d. 1266)
and known as Urd\hspace{-1.5 mm}$_.$\hspace{0.5 mm}\i\hspace{-1.5 mm}$^-$'s
Lemma, can be described as follows: ``Given any two equal
lines that form equal angles with a base line, either internally or externally,
the line joining the extremities of those two lines would be parallel to the
base line.''$^{33}$  

Another important astronomer was Ibn 
al-Sha\hspace{-2 mm}$^-$t\hspace{-1.0 mm}$_{.}$ir (d. 1375),
who was from Damascus.
He devised a model of the Moon's motion that was mathematically identical
to the one devised by Copernicus.  Victor Roberts's article on Ibn 
al-Sha\hspace{-2 mm}$^-$t\hspace{-1.0 mm}$_{.}$ir
is subtitled, ``A pre-Copernican Copernican model.''$^{34}$  

We now turn to two curious aspects of Copernicus's famous book.  Without proving
it, he made use of 
Urd\hspace{-1.5 mm}$_.$\hspace{0.5 mm}\i\hspace{-1.5 mm}$^-$'s
Lemma.  In Book 3, chapter IV, we find a curious diagram 
relating to the T\hspace{-2 mm}$_{.}$\hspace{1 mm}$\overline{\rm{u}}$s\i\hspace{-1.5 mm}$^-$
couple.  Various points are labeled in Roman capital 
letters.$^{35}$ Willy Hartner pointed out in 1973 that there is a similarity of 
Copernicus's diagram to 
Nas\hspace{-1.5 mm}$_{.}$\hspace{0.5 mm}\i\hspace{-1.5 mm}$^-$\hspace{-1 mm}r 
al-D\i\hspace{-1.5 mm}$^-$\hspace{-1 mm}n's
proof of 1260-61.$^{36}$  
Nas\hspace{-1.5 mm}$_{.}$\hspace{0.5 mm}\i\hspace{-1.5 mm}$^-$\hspace{-1 mm}r 
al-D\i\hspace{-1.5 mm}$^-$\hspace{-1 mm}n labels 
certain points {\em alif} (A), {\em baa} (B), {\em daal} (D), {\em heh} (H), {\em 
djim} (G), and {\em zaay} (Z).  Copernicus added an additional circle and uses two 
additional letters, and uses A, B, D, H, G, and F for the corresponding points in 
the diagram of Ibn 
al-Sha\hspace{-2 mm}$^-$t\hspace{-1. mm}$_{.}$ir.
It is elementary to conclude that Copernicus must 
have copied the diagram from somewhere.

So far as we know, none of the Arabic documents written by the astronomers of the 
``Maragha school'' (which includes Ibn 
al-Sha\hspace{-2 mm}$^-$t\hspace{-1.0 mm}$_{.}$ir)
were ever translated into 
Latin.  Yet here we are. Copernicus was born two hundred years after the death of 
Nas\hspace{-1.5 mm}$_{.}$\hspace{0.5 mm}\i\hspace{-1.5 mm}$^-$\hspace{-1 mm}r 
al-D\i\hspace{-1.5 mm}$^-$\hspace{-1 mm}n,
and one hundred years after the death of Ibn 
al-Sha\hspace{-2 mm}$^-$t\hspace{-1.0 mm}$_{.}$ir. Copernicus's 
book {\em De Revolutionibus Orbium Coelestium} uses the 
T\hspace{-2 mm}$_{.}$\hspace{1 mm}$\overline{\rm{u}}$s\i\hspace{-1.5 mm}$^-$
couple, Urd\hspace{-1.5 mm}$_.$\hspace{0.5 mm}\i\hspace{-1.5 mm}$^-$'s Lemma, 
and his model of the Moon's motion is mathematically equivalent to that of Ibn 
al-Sha\hspace{-2 mm}$^-$t\hspace{-1.0 mm}$_{.}$ir.
Ragep (2016) has recently argued that, ``Ibn 
al-Sha\hspace{-2 mm}$^-$t\hspace{-1.0 mm}$_{.}$ir's models in 
fact have a `heliocentric bias' that made them particularly suitable as a basis 
for the heliocentric and `quasi-homocentric' models found in the {\em 
Commentariolus}''$^{37}$ (the preliminary exposition of Copernicus's heliocentric 
model, from the year 1514). But it should be noted that Copernicus's placement of 
the Sun at the center of the solar system was his invention alone. It was not just 
a matter of reversing the direction of the Earth-Sun vector in a model of Ibn 
al-Sha\hspace{-2 mm}$^-$t\hspace{-1.0 mm}$_{.}$ir.$^{38}$

In his masterful translation of, and commentary on, Copernicus's {\em 
Commentariolus}, Noel Swerdlow notes four similarities between the work of Ibn 
al-Sha\hspace{-2 mm}$^-$t\hspace{-1.0 mm}$_{.}$ir and
Copernicus and writes, ``...the identity with the earlier planetary 
theory of Copernicus's models for the moon {\em and} the first anomaly of the 
planets {\em and} the variation of the radius of Mercury's orbit {\em and} the 
generation of rectilinear motion by two circular motions seems too remarkable a 
series of coincidences to admit the possibility of independent 
discovery.''$^{39}$

How might Copernicus have come to know aspects of medieval Arabic astronomy? 
Consider that in 1453 Constantinople fell to the Turks, and the Ottoman Empire was 
established.  Christian scholars and their documents could have been transported 
to northern Italy, where Copernicus did his ``graduate work'' in astronomy.  In 
1957 Otto Neugebauer discovered a Greek-language manuscript (which now resides in 
the Vatican library, and is catalogued as Vatic. gr. 211).  It contains a diagram of 
T\hspace{-2 mm}$_{.}$\hspace{1 mm}$\overline{\rm{u}}$s\i\hspace{-1.5 mm}$^-$'s
couple, but nothing related to 
al-Urd\hspace{-1.5 mm}$_.$\hspace{0.5 mm}\i\hspace{-1.5 mm}$^-$,$^{40}$
who was a contemporary.  No reference would be expected to Ibn 
al-Sha\hspace{-2 mm}$^-$t\hspace{-1.0 mm}$_{.}$ir
as Vatic. gr. 211 was written before 1308.$^{41}$ In any case, this accidental 
discovery by Neugebauer raised major expectations regarding new information on the 
transmission of Arabic astronomical knowledge from the Eastern Mediterranean
to Italy and beyond.

Ragep (2014) has recently written on the Persian astronomer Shams Pouchares and the transmission
of the T\hspace{-2 mm}$_{.}$\hspace{1 mm}$\overline{\rm{u}}$s\i\hspace{-1.5 mm}$^-$
couple from Tabriz to Byzantium and on to Italy through the works of Gregory
Chioniades.$^{42}$  Also,
thanks to the researches of Tzvi Langermann and Robert Morrison, we 
learn of the Jewish scholar Moses Galeano, who lived in Constantinople, Crete, and 
the Veneto (the region that contains Venice and, most notably, Padua).  To a 
significant degree he understood the astronomy of Ibn 
al-Sha\hspace{-2 mm}$^-$t\hspace{-1.0 mm}$_{.}$ir. It is highly 
likely that he passed on such knowledge to the astronomers in Padua in the years 
1497 to 1502, overlapping the time Copernicus was living there.$^{43,44}$
As Saliba points out toward the end of {\em Islamic Science and the Making of
the European Renaissance}, we are only beginning to understand the transmission of
astronomical knowledge from Asia Minor to Renaissance Europe.$^{45}$

% xxx

\newpage

\begin{center}
Acknowledgments
\end{center}

We thank Fernanda Strasser, Rebecca Shaftman, Shana Loshbaugh, and Owen Gingerich
for useful comments.  We also thank Jamil Ragep for useful comments and references.

\begin{center}
NOTES
\end{center}

\refs

1.  Valentin Rose, ``Ptolemaeus und die Schule von Toldeo,'' {\em Hermes}, 8 
(1874), 327-349.  In the preface to G. J. Toomer's English translation of 
Ptolemy's {\em Almagest} (p. vii), he points out, ``One 
can no longer assume that those with a serious interest in history are able to 
read German with ease.'' Rose's article is written in very difficult German, with 
many footnotes, primarily in Latin, and an appendix giving the beginning and end 
of the book by Daniel of Morley, entirely in Latin.  The German passages here have
been translated by one of us (KK).  There are two places noted where we were 
required to paraphrase.  The Latin passages presented here were translated by 
the second coauthor (BB).  

2.  The notion that a formal school of translators functioned in
medieval Toledo has been the subject of much debate. It can be traced to Amable
Jourdain's mention that a ``school of translators'' was created in this city by
Raymond of Toledo, who became Archbishop of this city c. 1124. See Amable Jourdain,
{\em Recherches critiques sur l’\^{a}ge et l'origine des traductions latines d'Aristote: et
sur des commentaires grecs ou arabes employ\'{e}s par les docteurs scolastiques} (Paris:
Joubert, 1843), pp. 108, 119. Rose's article has a fundamental place in this controversy,
since it offers the first search for evidence to support its existence.

3.  According to an on-line entry by Albert Frederick Pollard in the {\em Dictionary of 
National Biography}, 1885-1900, vol. 39, Daniel flourished from 1170 to 1190.  
His book is called both ``Philosophia Magistri Danielis de Merlac''and ``Liber 
de Naturis inferiorum et superiorum.''  See:
https://en.wikisource.org/wiki/Morley,\_Daniel\_of\_(DNB00)

4.  [Rose, p. 331, note 3:] The book begins with a general consideration 
concerning the creation of the world (of man in God's image) $-$ different worlds: 
{\em mundus archetypus} (used by Chalcidius [4th century translator of Plato from 
Greek to Latin] in [Plato's dialogue] {\em Timaeus} concerning yle 
([$\upsilon\lambda\eta$], {\em materia}) four elements, opinions concerning {\em 
maiores de rerum principiis}, the arrangement of the lower and upper world (called 
Mercurius books).  Creation: four {\em rerum genera} [fundamental elements] 
(matter, form, {\em compositio} and {\em compositum}).  The ordering of the 
elements in the world.  Book II, folio 95: {\em de superiorum constitutione 
secundum Arabes}.  Heavens, stars in the eighth sphere (after Alfraganus).  The 12 
[zodiacal] signs.  The constellations (according to the Arabs) and their virtues.  
The first representation of the Arabic scientific school, according to Gerard's 
teaching.[Note that ``ylem'' was a term used by George Gamow, Ralph Alpher, and 
Robert Herman in their Big Bang theory of the 1940's.  Ylem is the primordial 
material out of which everything is created.]

5.  Bel\'{e}n Bistu\'{e}, {\em Collaborative Translation and Multi-Version Texts 
in Early Modern Europe} (London and New York: Ashgate, 2013), 58-59.  On p. 61 
there is discussion of two translators working together at the same time vs. two 
translation instances carried out at different times.  Daniel's use of the present
participle form ({\em interpretante}) indicates that Gerard and Galib worked simultaneously. 
On p. 63 we learn that Roger Bacon had a low opinion of Gerard, Hermann, Michael Scot, and 
Alvredus Anglicus.  Apparently, Roger's opinion was that if someone needed a 
collaborative partner to render the Arabic into an intermediate language, the 
translator was {\em mediocre}.

6.  The {\em Almagest}'s given title was Mathematical Syntaxis, meaning mathematical 
compilation.  It became known as the ``greatest compilation'' or {\em al-Majisti}, 
hence {\em Almagest}.

7. Here Rose inserts a parenthetic expression: el-feludi, from which [we obtain] 
Pheludiensis and then wrongly rendered Pelusiensis, instead of el-keludi = \'{o} 
K$\lambda\alpha\upsilon\delta\iota$o$\zeta$, according to Reiske in Museum [der] 
Alterthumswissenschaften, vol. II (Berlin, 1808). [Philip Buttmann's article is 
``\"{U}ber den Ptolemaus in der Anthologie und den Klaudius Ptolemaus,'' Museum der 
Alterthumswissenschaften, vol. 2, 1810, pp. 455-506, according to Toomer's article on 
Ptolemy in the {\em Dictionary of Scientific Biography}.]  A Google search on 
``Pheludiensis'' leads the reader to two copies of ``Almagestu[m] Cl. Ptolemei 
Pheludiensis Alexandrini astronomo[rum] principis. Opus ingens ac nobile omnes 
celoru[m] motus continens'' in the Barchas and Newton Collections, respectively, at 
Stanford University.  They were published in 1515 in Venice by Petrus Liechtenstein.  
These are two copies of the first published edition of the {\em Almagest}.  The main text 
runs to 152 folios of size 31 cm, plus tables and diagrams.  In the 1890 Catalogue of 
the Crawford Library of the Royal Observatory, Edinburgh, there is an entry for the 
1515 Venice edition, confirming that the number of folios was 152.

8.  The year of this translation is 212 in the Muslim calendar, which ran from
April 2, 827, through March 21, 828, in the Christian calendar.  For a 
Christian/Muslim calendar converter
see www.oriold.uzh.ch/static/hegira.html (accessed May 19, 2017).

9. Baldassare Boncampagni, {\em Della vita e delle opere di Gherardo Cremonese 
traduttore del secolo duodecimo e di Gherardo da Sabbionetta astronomo del secolo 
decimoterzo} (Atti dell'Accademia Pontificia de' Nuovi Lincei IV, Rome, 1851).

10. [Rose p. 334, note 1:] The Prologue of Geber makes 
mention of Ptolemy's {\em Almagest} as well as books of Theodosius, Menelaus and then
Euclid.  At the conclusion of the book is the well known signature: ``Hunc librum
transtulit in Toleto magister Girardus cremonensis de arabico in latinum''
(see Bonc. p. 16 and 5) [book translated from Arabic into Latin in Toledo
by Gerard of Cremona].

11. The manuscript in question is in the Laurentiana library in Florence.  See: 
Charles H. Haskins, and Dean Putnam Lockwood, ``The Sicilian translators of the 
twelfth century and the first Latin version of Ptolemy's Almagest,'' {\em Harvard 
Studies in Classical Philology}, 21 (1910), 75-102. Haskins and Lockwood (p. 78, 
note 1) give the key passage, in Latin, that that manuscript is a transcription, 
made in 1175 by one Thaddeus of Hungary, of Gerard's Latin translation of the 
{\em Almagest}.  This is the generally accepted date for the translation, but to make a 
point, it is an upper limit.  Gerard could have finished his translation before 
1175.  Gerard went to Toledo particularly to work on the {\em Almagest}, and he arrived 
no later than 1144.  Did he work his way up to it and keep refining his 
translation for years? Lemay, ref. 15 below, made the point nearly 40 years ago 
that Gerard could have finished his translation some time {\em before} 1175.

12. This key biographical and bibliographical information can be found in the 
following references: Michael McVaugh, ``A list of translations made from Arabic 
into Latin in the twelfth century,'' in Edward Grant, ed., {\em A Source Book in 
Medieval Science} (Cambridge: Harvard University Press, 1974), 35-38; and Charles 
Burnett, ``The coherence of the Arabic-Latin translation program in Toledo in the 
twelfth century,'' {\em Science in Context}, 14 (2001), 249-288.

13.  [Rose, p. 335, note 3:]``in m\"{u}ndlichem Dictate'' = oral dictation. 

14.  Mar\'{i}a Rosa Menocal, {\em The Ornament of the World: how Muslims, Jews 
and Christians created a culture of tolerance in medieval Spain} (Boston, New 
York, and London: Little, Brown and Co., 2002), on p. 197.

15.  Richard Lemay, ``Gerard of Cremona,'' in {\em Dictionary of Scientific 
Biography, Suppl. I}, (Charles Scribner's Sons, 1978), vol. 15, 173-192. 
Lemay (his note 14) further states, with respect to the oral dictation method 
referenced in ref. 13 above: ``We doubt, however, that this collaboration 
with a native Spaniard was necessarily done `in m\"{u}ndlichem Dictate,' as 
Rose states, taking his example from Rudolf of Bruges.  This process clearly 
prevented the Latin collaborator from working directly with the Arabic text.  
Such a situation, if probable for Rudolf, who made very few translations, is 
quite unthinkable for Gerard of Cremona.''

16.  [Rose, p. 338, note 1:] ``Whoever needed an Arabic book went to Toledo.''
Rose then follows with a long-winded footnote, mostly in Latin, relating to 
one Marcus Toletanus, a translator in Toledo of unknown period of time (but 
sometime in the twelfth century).

17.  H. G. Farmer, ``Virgilius Cordubensis,,'' {\em The Journal of the
Royal Asiatic Society of Great Britain and Ireland}, No. 3 (1929), 599-603.
One can easily conclude that Virgilius Cordubensis was a fictional person.

18.  [Rose, p. 346, note 1]: For Hermann's activity as translator we have the 
known dates (thanks to Jourdain) of 1240 and 1256, from which we know for the 
Summa Alexandrinorum of the Nicomachean Ethics (Renan Av. 213) the date 1244 
from the Bodleian codex Canonic. lat. cl. 271 (ch. XV).  Where did he live? 
Roger saw him without doubt in Paris.  Where was he a bishop? In Spain? A 
Hermann, Bishop of Astorga, 1266-1272 (according to Gams), is the only 
non-Hispanic name at this time.  That concurs with ``adhuc vivt'' [still 
living] of Roger from the year 1271, and also Robert the Englishman remained 
Archdeacon of Pamplona (Jourdain, p. 102).  In his testament of 12 November 
1272, that Hermann issued decrees concerning estates in Palencia (Florez, Esp. 
Sagr. XVI, 242).  Did he come from Palencia? From the university? As one of 
the invited foreigners?

% xxx

19.  George Sarton, {\em Introduction to the History of Science} (Huntington, 
New York: Robert E. Krieger, 1975), vol. I, pp. 274, 423, 545, 562, and 565.  

20.  G. J. Toomer, {\em Ptolemy's Almagest} (New York, Berlin, Heidelberg, Tokyo: 
Springer-Verlag, 1984), pp. 2-3.  Toomer also mentions an earlier Arabic translation
of the {\em Almagest} carried out under the auspices of
al-Ma'mu\hspace{-2 mm}$^-$\hspace{-0.5 mm}n by al-H\hspace{-2 mm}$_{.}$\hspace{1 mm}asan 
ibn Quraysh.  Given that a second was commissioned later in reign of 
al-Ma'mu\hspace{-2 mm}$^-$\hspace{-0.5 mm}n, one might suppose that the first
was deemed unsatisfactory.

21. A preliminary comparison of different names given to the constellations in Gerard's 
translation suggests that at least he had access to al-Hajjaj's translation. Bistu\'{e}, 
ref. 5 above, pp. 72-73.

22. Although far less influential than Gerard of Cremona's
version from the Arabic, there was an earlier Latin version of the Almagest made
directly from the Greek at the Sicilian court of William I (c. 1160), by an anonymous
student who worked with the help of a Greek expert named Eugene. It survives in an
early-fourteenth-century copy that formed part of the library of Coluccio Salutati, a
testimony that this version was known in the humanist Florentine circle. The next
known version made directly from the Greek is that of George of Trebizond (1451);
this was the first {\em published} version in Latin (Venice, 1528) that was translated 
directly from Greek.  See Sarton, ref. 19 above, and Haskins and Lockwood, ref. 
11 above,  pp. 78, 84.

23.  Andr\'{e} Goddu, ``Nicholas Copernicus,'' in {\em Complete
Dictionary of Scientific Biography}, vol. 20 (Detroit: Charles Scribner's Sons, 2008),
176-182.

24.  C. Doris Hellmann and Noel M. Swerdlow, ``Georg Peurbach,'' in {\em Complete
Dictionary of Scientific Biography}, 15 (Detroit: Charles Scribner's Sons, 2008),
473-479.

25.  Edward Rosen, ``Johannes Regiomontanus,'' in {\em Complete
Dictionary of Scientific Biography}, vol. 11 (Detroit: Charles Scribner's Sons, 2008),
348-352.

26.  George Saliba, {\em Islamic Science and the Making of the European Renaissance}
(Cambridge, Massachusetts, and London: MIT Press, 2007), in particular pp. 17, 
151-165, 193-221, and 226-232.

27.  Kevin Krisciunas, ``Determining the eccentricity of the Moon's orbit 
without a telescope,'' {\em American J. of Physics}, 78 (2010), 828-833.

28. {\em Ibid.}, note 4.

29. E. S. Kennedy, ``Late Medieval Planetary Theory,'' {\em Isis}, 57 (1966), 365-378.

30. Saliba, ref. 26 above, p. 157.

31.  F. J. Ragep, 
Nas\hspace{-1.5 mm}$_{.}$\hspace{0.5 mm}\i\hspace{-1.5 mm}$^-$\hspace{-1 mm}r 
al-D\i\hspace{-1.5 mm}$^-$\hspace{-1 mm}n
al-T\hspace{-2 mm}$_{.}$\hspace{1 mm}$\overline{\rm{u}}$s\i\hspace{-1.5 mm}$^-$'s
{\em Memoir on Astronomy} (al-Tadhkira fi
`ilm al-hay'a), vol. I (New York, Berlin, Heidelberg: Spring-Verlag, 1993), 
in particular pp. 196-199.

32. https://en.wikipedia.org/wiki/Nasir\_al-Din\_al-Tusi (accessed June 23, 2017).

33.  Saliba, ref. 26 above, p. 152, and Fig. 6.4 on p. 203.

34.  Victor Roberts, ``The Solar and Lunar Theory of Ibn 
ash-Sha\hspace{-2 mm}$^-$t\hspace{-1.0 mm}$_{.}$ir: A Pre-Copernican
Copernican Model,'' {\em Isis}, 48 (1957), 428-432.

35.  A. M. Duncan, {\em Copernicus: On the Revolutions of the Heavenly Spheres.  
A New Translation from the Latin, with an introduction and notes}, (Newton
Abbot, London, and Vancouver: David \& Charles, 1976), pp. 145-146.

36.  Willy Hartner, ``Copernicus, the Man, the Work, and Its History,'' {\em Proc. Amer.
Phil. Soc.}, 117 (1973), 413-422.  A much clearer diagram lettered in Arabic is
found in Ragep, ref. 31 above, p. 199.

37.  F. Jamil Ragep, ``Ibn 
al-Sha\hspace{-2 mm}$^-$t\hspace{-1.0 mm}$_{.}$ir and
Copernicus: The Uppsala Notes Revisited,''
{\em J. for the Hist. of Astronomy}, 47 (2016), 395-415, on p. 395.

38.  Personal communication, Owen Gingerich to Kevin Krisciunas, 23 March 2017:
``I vehemently disagree that the Ibn al-Shatir geometry would have 
inspired a heliocentric arrangement.  On the other had, I have lately become sorry 
that I didn't explain in my recent Copernicus biography how it might have been 
very useful to Copernicus later on after he was exploring the heliocentric 
arrangement.  If he just took the Ptolemaic arrangements and stacked all the 
planetary apparatus around the Sun, there would be the huge and unseemly clutter 
of all the equants near the Sun.  Using the Ibn al-Shatir arrangements, each 
mechanism is associated closely with the individual planets.  Copernicus could 
have worked this out for himself once given a hint.  It may have saved the 
heliocentric transformation, but there is no good reason to think it might have 
engendered it. I can imagine Copernicus was very excited when he realized this 
move would tidy up his system, and he may never have heard of Ibn al-Shatir.''
In a personal communication of Jamil Ragep to Kevin Krisciunas of September 10, 2017,
Ragep comments, ``What has been overlooked is
that the virtual identity of Ibn al-Shatir's models to those of Copernicus is
not confined to the lunar model. In fact, as my student Sajjad
Nikfahm-Khubravan and I show in an article (currently under review for {\em JHA}),
the Mercury model of Ibn al-Shatir is identical to that of Copernicus's in $De$ 
$Rev$[$olutionibus$] and in fact the transformation to a heliocentric model is trivial. This
was already remarked on in the article by Kennedy and Roberts, but somehow
the implications have not been recognized, even by Swerdlow and Neugebauer.
As you know, this is the most complex of Copernicus's longitude models and it
seems to me implausible to think that Copernicus would come up with exactly
the same model when there were so many other possibilities.''

39.  Noel M. Swerdlow, ``The Derivation of the First Draft of Copernicus's Planetary
Theory: A Translation of the Commentariolus with Commentary,'' {\em Proc. Amer.
Phil. Soc.}, 117 (1973), 423-512, on p. 469.  See also p. 504.

40.  Saliba, ref. 26 above, p. 214. 

41.  E. A. Paschos and P. Sotiroudis, {\em The Schemata of the Stars: Byzantine
Astronomy from A. D. 1300}, (Singapore: World Scientific, 1998), p. 19.

42. F. J. Ragep, ``New Light on Shams: The Islamic Side of {\em Shams Pouchares}.''
In {\em Politics, Patronage, and the Transmission of Knowledge in 13th - 15th Century
Tabriz}, edited by Judith Pfeiffer, (Leiden: E. J. Brill, 2014), pp. 231-247.
See also {\em The Schemata of the Stars}, ref. 41 above, and Saliba, ref. 26 above, 
p. 194.

43. Y. Tzvi Langermann, ``A Compendium of Renaissance Science: {\em Ta`alumot 
hokmah} by Moses Galeano,'' {\em Aleph: Historical Studies in Science and 
Judaism}, 7 (2007), 283-318.

44.  Robert Morrison, ``A Scholarly Intermediary between the Ottoman Empire and Renaissance
Europe,'' {\em Isis}, 105 (2014), 32-57.

45. Saliba, ref. 26 above, pp. 210-217.

%At the end of Gerard's translation of Galen's [Greek physician, ca. 129-210 AD]
%{\em Tegni cum commento Haly} [Haly Abenrudian = Ali ibn Ridwan, ca. 988-1061]
%(``Tegni Galieni cum expositione ali abrodoan'', Index of Bonc., p. 6)
%his younger collaborators (``socii'') had the late master and his ``last works'' in mind 
%as is well known shortly after his death.  In imitation of 

%zzz

%Nas\hspace{-1.5 mm}$_{.}$\hspace{0.5 mm}\i\hspace{-1.5 mm}$^-$\hspace{-1 mm}r 
%al-D\i\hspace{-1.5 mm}$^-$\hspace{-1 mm}n
%al-T\hspace{-2 mm}$_{.}$\hspace{1 mm}$\overline{\rm{u}}$s\i\hspace{-1.5 mm}$^-$

%Mu'ayyad al-D\i\hspace{-1.5 mm}$^-$\hspace{-1 mm}n
%al-Urd\hspace{-1.5 mm}$_.$\hspace{0.5 mm}\i\hspace{-1.5 mm}$^-$

%Sha\hspace{-2 mm}$^-$t\hspace{-1.0 mm}$_{.}$ir

%al-H\hspace{-2 mm}$_{.}$\hspace{1 mm}ajja\hspace{-2 mm}$^-$j ibn Yu\hspace{-2 mm}$^-$suf

%Ish\hspace{-1.5 mm}$_.$\hspace{0.5 mm}a\hspace{-2 mm}$^-$\hspace{-0.5 mm}q 
%ibn H\hspace{-2 mm}$_{.}$\hspace{1 mm}unayn

%Tha\hspace{-2 mm}$^-$\hspace{-0.5 mm}bit

%Ma'mu\hspace{-2 mm}$^-$\hspace{-0.5 mm}n

%Nas\hspace{-2mm}$_{.}\overline{\rm{\i}}$r al-D$\overline{\rm{i}}$n
%al-T\hspace{-2 mm}$_{.}$\hspace{1 mm}$\overline{\rm{u}}$s$\overline{\rm{i}}$

%H\hspace{-2 mm}$_{.}$\hspace{1 mm}ello.

%al-H\hspace{-2 mm}$_{.}$\hspace{1 mm}asan ibn Quraysh

%\refs

%Krisciunas, K., Phillips, M. M., \& Suntzeff, N. B., 2004, ``Hubble Diagrams of 
%Type Ia Supernovae in the Near-Infrared,'' {\em Astrophys. J.}, {\bf 602},
%pp. L81-L84.

\end{document}